\documentstyle[sprocl]{article}

\def\mxth{\mathsurround=0pt }
\def\xversim#1#2{\lower2.pt\vbox{\baselineskip0pt \lineskip-.5pt
  \ialign{$\mxth#1\hfil##\hfil$\crcr#2\crcr\sim\crcr}}}             
                                    \def\lesssim{\mathrel{\mathpalette\xversim <}}                                    
\def\Journal#1#2#3#4{{#1} {\bf #2}, #3 (#4)}

\def\NPB{{\em Nucl. Phys.} B}
\def\PLB{{\em Phys. Lett.}  B}
\def\PRL{\em Phys. Rev. Lett.}

\def\ra{\rightarrow}
\def\be{\begin{equation}}
\def\ee{\end{equation}}
\def\bea{\begin{eqnarray}}
\def\eea{\end{eqnarray}}

\begin{document}

\begin{titlepage}
\noindent

\phantom{a}     \hfill         
RU--98--31              \\
\phantom{a}     \hfill         
hep-ph/9809422          \\[3ex] 
\begin{center}

{\bf SUPERGRAVITY MIRACLES:\\
PHENOMENOLOGY WITH GRAVITATIONAL DIVERGENCES}\\[5ex]

{NIR POLONSKY}\\[1ex]

{\it Physics Department, Rutgers University, Piscataway, NJ 08854-8019, USA
\\E-mail: nirp@physics.rutgers.edu
} 

\end{center}
\vspace{1cm}

{\begin{center} ABSTRACT \end{center}}
\vspace*{1mm}
{\noindent

The properties of singlet fields in supergravity background are discussed.
Particularly, 
it is shown that the ${\cal{O}}(\Lambda^{2})$ contributions
to their one-point function 
lead to novel derivations of 
effective scales such as the $\mu$ parameter, a messenger scale, 
an intermediate symmetry-breaking scale, and the unification scale,
in terms of the scale of spontaneous supersymmetry-breaking in a hidden sector.
The singlet one-point function effectively serves 
as a messenger of supersymmetry breaking.
}

\vspace*{2truecm}
{\begin{center} 
{\it Talk presented at PASCOS-98} \\
{\it The Sixth International Symposium  on Particles,  Strings and Cosmology}\\
{\it Northeastern University, Boston, MA, March 22 -- 29, 1998}
\end{center}}

\vfill
\end{titlepage}

If supersymmetry is realized at sub-TeV energies then the effective 
Lagrangian contains many new parameters which explicitly but softly 
break supersymmetry in the Standard Model (SM) sector
(i.e., only new logarithmic divergences are introduced).
The soft supersymmetry breaking (SSB) parameters 
lift the mass degeneracy between the ordinary matter and the s-particles.
Consistent high-energy constructions  usually contain a hidden sector --
with {\it no tree-level renormalizable interactions} with the SM
(observable) sector -- in which supersymmetry is broken spontaneously
at a scale $M_{SUSY}$. Some other forms of hidden-observable interactions 
then mediate the SSB parameters in the observable sector.
The absence  of large contributions from s-particle loops to 
flavor changing neutral currents implies that unless the 
SSB parameters are in the tens of TeV range  they cannot be
arbitrary: The s-fermion mass-squared matrices have to be diagonal
to high accuracy in the same basis as the fermion mass matrices.
This is the well-known flavor puzzle. Its solutions
lead to only a small number of families of viable models 
for the mediation of the supersymmetry breaking from the hidden
to the observable sector. In addition,
the absence of an access of  $W^{+}W^{-}$-like events
in recent LEP runs implies that the Higgs fermions (Higgsinos)
must be massive with a mass $\mu \sim {\cal{O}}$(0.1 - 1 TeV). 
Hence, one has to introduce a common (supersymmetric) mass for the Higgsinos
and Higgs bosons, $\mu H_{1}H_{2}$, in the low-energy superpotential.
The $\mu$-parameter has to be roughly of the order of the SSB
parameters even though it does not break the global supersymmetry
at low-energies, hinting that its origins are similar
to those of the SSB parameters. This is the well-known $\mu$-puzzle.
Both puzzles provide  crucial hints in deciphering the high-energy
theory and will be explored below from a new perspective.

If {\it tree-level  non-renormalizable supergravity 
interactions} are the mediator of supersymmetry
breaking then one typically assumes $m_{SSB} \sim
(M_{SUSY}^{2}/M_{P})$,
hence fixing $M_{SUSY} \sim 10^{10 - 11}$ GeV.
$\mu \sim M_{SUSY}^{2}/M_{P}\sim m_{SSB}$ can naturally arise in this case.  
However, in order to (technically) solve the flavor puzzle,
the field-space (Kahler) metric, which determines 
the flavor structure of $m_{SSB}^{2}$, has to
be proportional to the identity  
at least in each $ 3 \times 3$ flavor space. 
Such scenarios are often labeled as supergravity mediation.
Here, we will show that at the quantum level supergravity can also  induce 
${\cal{O}}(M_{SUSY}^{4}/M_{P})$ and ${\cal{O}}(M_{SUSY}^{2})$ 
scales in the observable sector~\cite{papers};
extending the meaning of supergravity mediation.
Small ${\cal{O}}(m_{SSB})$ parameters can either be 
${\cal{O}}(M_{SUSY}^{2}/M_{P})$ as before or, instead,
can be non-trivial functions of the larger ``new'' scales.
The above observation leads to
new mechanisms for $\mu \sim m_{SSB}$ generation; for inducing intermediate
(messenger or symmetry breaking) scales; and surprisingly, for relating
the unification scale to the Planck scale by a loop factor.

The new scales 
appear if the observable sector contains a singlet
chiral superfield $S = s + \theta \Psi_{s} +\theta^{2} F_{s}$.
$S$ may carry non-trivial charges under
global phase or discrete symmetries
of the superpotential, which 
provide selection rules for its couplings.
However, such symmetries are likely to be broken 
explicitly by non-holomorphic operators in the Kahler potential $K$
and/or once supersymmetry is broken in a hidden sector.
In this case, the singlet does not carry any 
conserved charges and it develops, due to quadratically divergent
quantum supergravity effects,
linear potential terms in its physical ($s$) and auxiliary ($F_{s}$)
scalar components which shift its potential~\cite{new}:
$V \rightarrow V + [\gamma(M_{SUSY}^{4}/M_{P})s +\beta \epsilon
M_{SUSY}^{2}F_{s} + h.c.]$.
(Note that the second term can appear independently in the case
of superpotential light-heavy mixing~\cite{old}.) 
The parameters $\gamma$ and $\beta 
\sim N/(16\pi^2)^{n}$ are  products of unknown loop
factors and  of counting factors $N$ which sum all relevant loops.
They are typically of the same order $\gamma,\,\beta \sim 10^{-2
\pm2}$ (assuming perturbativity, i.e.,
$N \lesssim 100$, and no accidental cancelations beyond one-loop,
$n \leq 2$), but can carry independent phases.
The parameter $\epsilon$ is a measure in Planckian units 
of the expectation values $\langle z \rangle$ of the supersymmetry
breaking fields in the hidden sector
(i.e., $\langle F_{z} \rangle \simeq M_{SUSY}^{2}$). 
The shifts, which explicitly break the supersymmetry in the $S$ sector, 
appropriately vanish in the supersymmetric limit, $M_{SUSY} \rightarrow 0$. 
However, they vanish more slowly then the conventional terms
$\sim M_{SUSY}^{2}/M_{P}$, and could
have important implications, in particular,
in the case of a low-energy $M_{SUSY} \sim 10^{5} - 10^{8}$ GeV. 
Specific applications are discussed below
in parallel to the classification of the possible effective  scales.
It can be most conveniently done, as outlined below, 
by considering the stabilization of the scalar potential.
For detailed analysis and discussion  and for complete references 
see Ref.~1, which we follow.

Consider the  superpotential terms $W = (\kappa/3)S^{3} + 
\lambda S\Phi_{1}\Phi_{2}$ (for $\kappa/2 < \lambda$),
leading to a bounded potential
$V(s) \sim [(M_{SUSUY}^{4}/M_{P})s + h.c.] +  \kappa^{2} s^{4}$.
Taking $\epsilon \rightarrow 0$ and neglecting all 
${\cal{O}}(M_{SUSY}^{2}/M_{P})$ terms, 
it is straightforward to show that
$\langle s \rangle  \sim (M_{SUSY}^{4}/\kappa^{2} M_{P})^{1/3}$ 
and $\langle F_{s} \rangle \sim \kappa\langle s \rangle ^{2}$.
Upon integrating out $S$
one finds a supersymmetric mass parameter $ W \sim \lambda \langle s
\rangle\Phi_{1}\Phi_{2}$ in the effective superpotential,
and  a SSB mass term $\sim \lambda \langle F_{s} \rangle 
\phi_{1}\phi_{2} + h.c.$ in the scalar potential (which lifts the
otherwise mass degeneracy between the scalar and fermion
components of the $\Phi$-fields).
One can identify $\Phi_{1}\Phi_{2} \ra H_{1}H_{2}$, 
generating the correct supersymmetric and SSB
mass terms in the Higgs sector in the case
of low-energy supersymmetry breaking $M_{SUSY} \sim 10^{6 \pm 1}$ GeV,
e.g.,  in the  framework of gauge-mediation (GM).
This is a unique supergravity solution to the $\mu$-puzzle in this
case  where  otherwise 
$\mu \sim M_{SUSY}^{2}/M_{P} \ll $ GeV is negligible.

Alternatively, one could identify
$\Phi_{1}\Phi_{2}$ with  messenger fields of GM whose
{\it gauge interactions} with the SM gauge superfields
are responsible for the generation of the usual 
SSB parameters {\it at the quantum level}.
Generally, one has in these models the flavor-universal 
loop relation $m_{SSB} \sim (\alpha/4\pi)\langle F_{s}/s\rangle$
(resolving the flavor puzzle without any assumptions with regard 
to the Kahler metric). This leads to the constraint 
$(4\pi/\alpha)M_{W} \sim \langle F_{s}/s\rangle \sim \langle s \rangle$ where
$\alpha = g^{2}/4\pi$ is a SM gauge coupling,
assuming that the messenger fields are in non-trivial
representations of the SM (e.g., 5 and $\bar{5}$ of SU(5)).
This identification constrains 
$M_{SUSY} \sim 10^{8}$ GeV and leads to a 
most minimal supergravity-triggered gauge-mediated model
which does not have any carefully constructed sector 
(as in typical GM models) to mediate between the hidden sector 
and the messenger fields $S$, $\Phi_{1}$ and $\Phi_{2}$. 
(The above resolution of the $\mu$ puzzle does not apply in this case.)
Finally, the messenger fields
may carry only (gauge) non-SM horizontal charges. If $M_{SUSY} \sim 10^{10}$
GeV, horizontal GM
can render those SM fields which carry horizontal
charges heavy, in the tens of TeV range -- realizing the
decoupling solutions to the flavor puzzle --  while
the SSB parameters, e.g., in the horizontally neutral Higgs sector, are 
$\sim M_{SUSY}^{2}/M_{P} \sim M_{W}$ as usual.

Turning off the singlet self  interaction ($\kappa = 0$)
the potential could be stabilized instead 
in a more subtle way by turning on $\epsilon \simeq 1$. 
One has $V \sim \gamma(M_{SUSY}^{4}/M_{P})s + h.c
+ \sum_{i}|F_{i}|^{2}$ and $F_{s} \sim \lambda\phi_{1}\phi_{2}
-\beta M_{SUSY}^{2}$.  V is then minimized 
(along a flat $D$ and $F$ direction) for 
$\langle \phi_{1} \rangle = \langle \phi_{2} \rangle =
\sqrt{\beta/\lambda}M_{SUSY}$, providing the seed for intermediate-scale
breaking of an extended symmetry (and possibly for see-saw neutrino masses)
if the $\Phi$ fields are identified with the Higgs bosons of that symmetry.
A most interesting situation arises if one couples the
singlet  to the the usual and extended symmetry Higgs fields simultaneously, 
$W = \lambda S\Phi_{1}\Phi_{2}+  \lambda^{\prime} S H_{1}H_{2}$,
with $\lambda > \lambda^{\prime}$. The latter Yukawa 
coupling hierarchy determines $\langle \phi \rangle \simeq M_{SUSY}$,
and $V(s) \sim \gamma(M_{SUSY}^{4}/M_{P})s + \langle \phi
\rangle^{2}s^{2}$ leads to stabilization due to the effective mass
term,  and henceforth to $\langle h \rangle \sim
\langle s \rangle \sim M_{SUSY}^{2}/M_{P}$. 
It provides an unconventional solution
to the $\mu$ puzzle in the usual supergravity picture 
($M_{SUSY} \sim 10^{10}$ GeV). 

There could also be a situation in which $\kappa = 0$ and
$\epsilon = 0$. (The case where both are non-zero
is more complicated and does not lead
to any interesting new results.) In this case,
$V(s) = (M_{SUSY}^{2}/M_{P})s^{2} - \gamma (M_{SUSY}^{4}/M_{P})s$
is minimized for $\langle s \rangle \simeq \gamma M_{P}$,
thus generating a scale separated from the Planck scale $M_{P}$
by only a loop factor. (Note that $M_{SUSY}^{2}/M_{P}$ terms are not
negligible here because of the otherwise unboundedness of $V(s)$.)
If we identify the $\Phi$ fields
with (two distinctive) self-adjoint fields of a grand-unified 
group then $\langle s \rangle \simeq \gamma M_{P}$ provides
the seed for the breaking of the unified symmetry at
the correct scale. ($K$ is constrained so that $N$ is not too large.)
In fact, a more careful examination shows that 
since $\langle F_{s} \rangle \sim \langle \phi_{1}\phi_{2} \rangle
\sim {\cal{O}}(\gamma M_{SUSY}^{2})$ then
$\epsilon \neq 0$ does not alter the above results. 
Hence, in models with two singlets, $S$ and $X$, which carry
different non-conserved global charges, the unification scale
can be realized simultaneously with an intermediate scale
(either a messenger scale for $\kappa_{X} \neq 0, \, \epsilon = 0$, or an
extended symmetry breaking scale for $\kappa_{i} = 0, \, \epsilon \neq 0$).

It is interesting to note that by fixing $m_{SSB} \sim
M_{SUSY}^{2}/M_{P}\Rightarrow M_{SUSY} \simeq 10^{10}$ GeV
and identifying  $\Phi_{1}\Phi_{2} \ra H_{1}H_{2}$, one always destabilizes 
the SM Higgs fields to scales far above the weak scale~\cite{old,new}.
While this is an undesirable special case, the above discussion
clearly indicates that it is a result of the assumptions made rather 
than of the presence of a true low-energy singlet in the theory~\cite{papers}.
On the contrary, the singlet is more generically a useful (and often
ignored) model building tool which enables  unique realizations
of low and intermediate scales in terms of the scale
of spontaneous supersymmetry breaking in a hidden sector.
Lastly, we note in passing that divergent contributions
to the two-point function of a light field also
arise from  supergravity quantum effect. This is most easily seen
by replacing (in  $K$ or in $V$) the singlet S with the light-field operator 
$QQ^{\dagger}/M_{P}$. One then finds corrections to the SSB
squared masses $\sim \gamma (M_{SUSY}^{2}/M_{P})^{2}$. Clearly, 
such corrections are relevant for the discussion of the flavor puzzle
as they effectively correct any flavor universal Kahler metric at 
the quantum level in supergravity. They are discussed elsewhere~\cite{papers}.

It is pleasure to thank C.~Kolda, H.~P.~Nilles, and
S.~Pokorski, for their collaboration.
Work supported by NSF grant No.~PHY-94-23002.

\end{document}